\documentstyle[aps,epsf]{revtex}
\def \be   {\begin{equation}}
\def \ee   {\end{equation}}
\def \l {\label}
\begin{document}
\input epsf
\baselineskip=25pt
\title{Dynamics and causality constraints}
\author{Manoelito M de Souza}
\address{{Departamento de
F\'{\i}sica - Universidade Federal do Esp\'{\i}rito Santo\\29065.900 -Vit\'oria-ES-Brazil}\thanks{ E-mail: manoelit@cce.ufes.br}}
\date{\today}
\maketitle
\begin{abstract}
\noindent The physical meaning and the geometrical interpretation of  causality implementation in classical field theories are discussed. Causality in field theory are kinematical constraints dynamically implemented via solutions of the field equations, but in a limit of zero-distance from the field sources part of these constraints carries a dynamical content that explains old problems of classical electrodynamics away with deep implications to the nature of physical interactions.
\end{abstract}
\begin{center}
\end{center}
\section{Introduction}
Causality implementation in field theory is naturally connected to the very concept of field propagation. Old and well known problems appear with a field in a close neighbourhood to its sources. Then a careful analysis is required as the kinematical constraint of a causal propagation is  mixed with the dynamics of the field-source interaction. In particular, 
for a point-like source, there are problems with infinities and other signs of inconsistency. Thus there is a generalized belief that these infinities are consequences of the source point-size dimension and, consequently, that a point-particle cannot be regarded as a viable model for a charged elementary physical object. This, as shown in [1], does not correspond to  reality. The infinities associated to a point-charge self-field are consequences of the way causality has being implemented with the use of lightcones, whose vertex is a singular point; the field infinity just reflects this singularity. This work returns to the ideas raised in [1] further discussing its  physical and  geometrical meanings. Although it is being based on the case of a point electric charge, its conclusions are of a wider generality, being valid not only for classical field theory but for any theory of fields defined with support on a conic hypersurface (Quantum Field Theory, Quantum Mechanics, General Relativity, Statistical Mechanics, etc), i.e. any theoretic frame work with causality implementation. It shows that the plain Maxwell's theory, in a short-distance limit, reveals unequivocal and previously unsuspected signs of a quantum nature (the existence of photons) through the indication of a discreteness on the electromagnetic interaction, hidden behind the classical continuous formalism. It hints a proposal of a new approach (fully developed elsewhere \cite{hep-th/0006237} where fields and sources are symmetrically treated as discrete objects from which the standard continuous fields are retreated as spacetime effective averages.

This paper is organized in the following way. The geometric vision of causality is discussed in Section II with the introduction of the new concept of extended causality in contraposition to the usual local causality and of their connection to wave-particle duality. Their relevance to point-charge electrodynamics and their inherent conflicts are discussed in Section III. The implications of extended causality on  field-source dynamics is exposed in Sections IV and V. The paper ends with some final comments and the conclusions in Section VI. 

\section{Causality and spacetime geometry}

The notation used is of omitting the spacetime indices when this causes no ambiguity. For example, $\partial$ for $\partial_{\mu}$, and $A(x,\tau)$ for a vector field $A^{\mu}(x,\tau)$;  $x$ stands for both, the event parameterized by $x^{\mu}=(t,{\vec x})$ and for the coordinate $x^{\mu}$ itself.

The propagation of a massless field on a flat spacetime of metric  $\eta_{\mu\nu}=diag(-1,1,1,1),$ is restricted by
\be
\label{0}
\Delta x^{2}=0,
\ee 
which defines a local double (past and future) lightcone: $\Delta t=\pm|\Delta{\vec x}|.$ This is also a mathematical expression of local causality in the sense that it is a restriction for the massless field to remain on this lightcone. It is a particular case of the more generic expression 
\be
\label{l}
\Delta\tau^2=-\Delta x^{2},
\ee which is, besides, the definition of the proper time $\tau$ associated to the propagation of a free physical object across $\Delta x$. 
As $\tau$ is a real valued parameter, the eq. (\ref{l}) just expresses that $\Delta x$ cannot be space like. Geometrically it is also the definition of a three-dimensional double cone, of which the lightcone and the time axis are  just the two extreme limiting cases. $\Delta x$ is the four-vector separation between a generic event $x$ and the hypercone vertex. This conic hypersurface is the support for the definition of a propagating field. 
The hypercone aperture-angle $\theta,\;\;0\le\theta\le\pi/4,$ is given by
$\tan\theta=\frac{|\Delta {\vec x}|}{|\Delta t|},\; c=1,
$
or $\Delta\tau^{2}=(\Delta t)^{2}(1-\tan^{2}\theta).$
A change of the supporting hypercone corresponds to a change of speed of propagation and is an indication of interaction.\\
On requirements of continuity one must consider the constraint (\ref{l}) on a neighbourhood of $x$: $(\Delta\tau+d\tau)^2=-(\Delta x+dx)^2$ or, after using eq. (\ref{l}), $\Delta\tau d\tau+\Delta x.dx=0$, which may be written as
\be
\l{tfdx}
d\tau+f.dx=0,
\ee
where $f$ is a four-vector tangent to the hypercone (\ref{l}). For $\Delta\tau\not=0$ it is just 
\be
\l{f}
f^{\mu}:=\frac{\Delta x^{\mu}}{\Delta\tau}{\Big|}_{\Delta\tau^2+\Delta x^2=0}
\ee
For $\Delta\tau=0$ the hypercone  (\ref{l}) reduces to the lightcone (\ref{0}) and $f$ to its tangent four-vector; $f$ and $\Delta x$ are both lightlike. It is important to observe that $f$ is well defined for any $\Delta\tau$, including $\Delta\tau=0$, as long as $\Delta x\not=0.$ A tangent is not defined at the hypercone vertex. This is a crux point, neglected in the existing literature \cite{Rorhlich,Jackson,Parrot,Teitelboim,Rowe,Lozada} which leads to the old and well known vexing problems of consistency in classical elctrodynamics \cite{hep-th/9610028}. Geometrically eq. (\ref{tfdx}) defines a hyperplane tangent to the hypercone (\ref{l}). The simultaneous imposition of eqs. (\ref{l}) and  (\ref{tfdx}) on the propagation of a free point object  produces a  much more stringent constraint than local causality as the object is restricted to remain on the intersection of the hypercone (\ref{l}) with its tangent hyperplane (\ref{tfdx}), that is, on the hypercone generator tangent to $f$, or the $f$-generator, for short. This corresponds to an extended concept of causality which will be referred as extended causality. 

Local and extended causality correspond to two distinct and complementary (like geometric and wave optics) description  of a same physical system. They correspond to different perceptions of the spacetime available to the free evolution of a physical system from a given initial condition, respectively as foliations of  hypercones and as congruences of straight lines, the hypercone generators. So, whereas the first one is appropriated for a description in terms of continuous and extended objects like a fluid, a field, a wave, the second one implies on a perception of them as sets of points, describing individually each point. Failure on recognizing this leads to problems of consistency in field theory at short distance.

\section{Point-charge electrodynamics}

Consider, for example, $z(\tau)$, the worldline of a classical point electron parameterized by its proper time $\tau;$ each event on this worldline belongs to  the (instantaneous) hypercone $$\Delta\tau^2+\Delta z^2=0$$  and the four-vector $u=\frac{dz}{d\tau}$ is tangent to the worldline (and to the hypercone). It satisfies $$d\tau+u.dz=0,$$ which corresponds to eq. (\ref{tfdx}). A free electron remains on the $u$-generator of its hypercone; an accelerated electron is  on a $u$-generator of its instantaneous hypercone. So, in a way, classical electrodynamics already uses extended causality for specifying the state of the classical electron, and this is consistent with an electron modeled as a point particle. This work discusses how extended causality is also used in the definition of its electromagnetic field and the conflicting problems brought with this, pointing the way to a new consistent formalism for field theory. \\
Consider now the electromagnetic field at $x$, emitted by this electron.   $\Delta x=x-z(\tau)$ defines a family of four-vectors connecting the event $x$ to events on the electron worldline $z(\tau).$ Then, accounting for the masslessness of the electromagnetic field, $\Delta x^2=0$, the double lightcone with vertex at $x,$ intercepts $z(\tau)$ at two points: $z(\tau_{ret})$ and $z(\tau_{adv}).$ See  the Figure 1. 
\hglue2cm

\begin{figure}
\vglue-5cm
\epsfxsize=400pt

\hglue2cm
\epsfbox{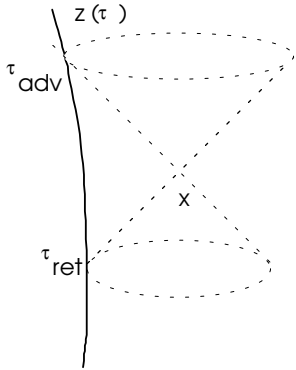}
\vglue-6cm
\caption{
The advanced and the retarded Li\`enard-Wiechert fields at an event x. $\tau_{adv}$ and $\tau_{ret}$ are the two intersections of the double hypercone $\Delta x^{2}=0$, for $\Delta x=x-z(\tau)$, with the electron worldline $z(\tau)$.}
\end{figure}  
 
The retarded field emitted by the electron at $z(\tau_{ret})$ must remain in the $z(\tau_{ret})$-future-lightcone, which contains x; and according to the standard interpretation \cite{Rorhlich,Jackson,Parrot}, the advanced field produced by the electron at $z(\tau_{adv})$ must remain in the $z(\tau_{adv})$-past-lightcone, which also contains x. So, the electromagnetic field is defined just with local causality. There is then supposedly a clear  dichotomy with respect to causality implementation in the treatment done to the electron and to its self-field \cite{BJP}, caused by the perception of the electron as a point particle, a discrete object, and of its field as a continuous and distributed one. Extended causality requires and implies discrete objects. 

The (retarded and advanced) Li\`enard-Wiechert solutions of classical electrodynamics \cite{Rorhlich,Jackson,Parrot,Teitelboim,Rowe,Lozada} are
\be
\l{aA}
A^{\mu}(x)=\frac{eu^{\mu}(\tau)}{\rho}{\bigg|}_{\tau=\tau_{s}},\qquad\hbox{ for}\quad \rho\not=0,
\ee
where $\tau_{s}$ stands for either $\tau_{ret}$ or $\tau_{adv}$, which are, respectively, the retarded and the advanced solution to the constraint 
\be
\l{xz}
(x-z(\tau))^2=0,
\ee
imposed to $A(x),$  and  $$\rho:=-u.\Delta x,$$ with $\Delta x==x-z(\tau)$, represents $|\Delta{\vec x}|$ in the charge rest-frame.  Although $A(x)$ is restricted just by eq. (\ref{0}), having thereby support on the lightcone,  for the calculation of its Maxwell field,
\be
\l{rotA}
{F_{\mu\nu}}:=\partial_{\nu}A_{\mu}-\partial_{\mu}A_{\nu},
\ee
on a point $x$ it is necessary to consider $A(x)$ on a neighbourhood of $x$, and so a constraint equivalent to the eq. (\ref{tfdx}) must be also considered to assure the consistency of eq. (\ref{0}) in this neighbourhood. From eq. (\ref{xz})  one has  $$\Delta x.d(x-z)=\Delta x.(dx-ud\tau)=0,$$ that allows to write
\be
\l{dlc}
d\tau+K.dx=0,
\ee
where K defined by
\be
\l{K}
K^{\mu}=\frac{\Delta x^{\mu}}{-u.\Delta x}{\Big|}_{\tau_{s}}=\frac{\Delta x^{\mu}}{\rho}{\Big|}_{\tau_{s}},
\ee
is a null $(K^{2}=0)$ four-vector, tangent to the lightcone $\Delta x^2=0$. $K^{\mu}$ shows the local direction of propagation of the electromagnetic field  emitted by the electron at $\tau_{s}.$  
In this context, eq. (\ref{dlc}) is a consistency relation of eq.  (\ref{l}) assuring its  validity for all successive pair of events $(x,z(\tau)).$
It implies on
 \be
K_{\mu}=-\frac{\partial\tau_{s}}{\partial x^{\mu}},
\ee
and then, 
\begin{equation}
\label{nablaA}
-\frac{1}{e}\partial_{\mu}A^{\nu}{\bigg|}_{\tau_{s}}=\bigg{\{}\frac{K_{\mu}a^{\nu}}{\rho}+\frac{u^{\nu}}{\rho^{2}}\partial_{\mu}\rho\bigg{\}}{\bigg|}_{\tau_{s}}=\frac{1}{\rho^{2}}{\bigg(}K_{\mu}W^{\nu}+u_{\mu}u^{\nu}{\bigg)}{\bigg|}_{\tau_{s}},
\end{equation}
where
$a:=\frac{du}{d\tau},$ 
\be
\l{delp}
\partial_{\mu}\rho{\bigg|}_{\tau_{s}}={\bigg\{}K_{\mu}{\bigg(}1+\rho a.{K}{\bigg)}-u_{\mu}{\bigg\}}{\bigg|}_{\tau_{s}}
\ee
and the ancillary four-vector function $W$,
\begin{equation}
\label{W}  
W^{\mu}=\{\rho a^{\mu}+u^{\mu}{\bigg(}1+\rho a.{K}{\bigg)}\}{\bigg|}_{\tau_{s}},
\end{equation}
 has been introduced just for notation simplicity. So, 
\be
\label{F} 
F^{\mu\nu}=\frac{1}{\rho^{2}}{\bigg(}K^{\mu}W^{\nu}-K^{\nu}W^{\mu}{\bigg)}{\bigg|}_{\tau_{s}}
\end{equation}
Geometrically the eq. (\ref{dlc}), like the eq. (\ref{tfdx}), defines a family of hyperplanes that, for $d\tau=0$, are tangent to the lightcone (\ref{xz}), and parameterized by $K_{\mu}=\eta_{\mu\nu}K^{\nu}.$  The use of both constraints (\ref{xz}) and (\ref{dlc}) implies the extended causality in the $F$ definition, exhibited on its explicit dependence on $K^{\mu}$, a four-vector tangent to a light-cone generator. But rigourously this is an inconsistent procedure as an undue mixing of local and extended causality on a same physical object. The inconsistency is on $F$ being defined as the curl of $A(x)$ which is a continuous field with support on the lightcone. If its support is reduced to the $K$-generator of its lightcone, $F$ should be regarded as a discrete object, similar, in this aspect, to its very source, the point electron. The problem, of course, is not with the definition  (\ref{rotA}) of $F$ but with $A(x)$ being a propagating field and, therefore, restricted by a causality constraint that necessarily requires the constraint (\ref{dlc}) on any field derivative. In other words, there would be no problem with the definition (\ref{rotA}) if $A(x)$ where not constrained  by (\ref{xz}) as the constraint (\ref{dlc}) would not be called up then\footnote{A completely consistent formulation would require $A(x)$ being defined with extended causality too. This would imply the consideration of fields defined with support on (1+1) sub manifolds imbedded in the (3+1) spacetime, ``discrete fields", with a complete symmetry between fields and sources, both being discrete objects. This is done in \cite{hep-th/0006237,hep-th/9911233}. The goal of the present paper is of pointing the existence of two modes for causality implementation (local and extended) in field theory and their implications to the meaning and nature of the fields and their interactions.} 

In the standard literature, without knowledge of extended causality, $F$ is seen as a field with support on the lightcone, i.e. a continuosly extended object with old and well-known problems with infinities and other inconsistencies. They just vanish after due consideration of extended causality \cite{hep-th/9610028}.

The origin of this imbroglio is that the equation (\ref{dlc}), as it can be formally obtained from a derivation of eq. (\ref{l}),  has  been historically considered [2-7] as if all its effects were already described by eq. (\ref{l}), included in it  and not, as it is the case, a new and independent restriction to be considered at a same footing and in addition to it. An evidence of this is that eqs. (\ref{l}) and (\ref{dlc}) carry distinct physical informations that will be discussed now. 

\section{Dynamics and causality}

Eq. (\ref{dlc}) connects the restriction on the propagation of the charge to the restriction on the propagation of its emitted or absorbed fields.  Like its parent equation (\ref{l}) it is just a kinematical restriction. But in the short-distance limit, when $x$ tends to $z(\tau)$, eq. (\ref{dlc}), in contradistinction to  eq. (\ref{l}),  is directly related to the changes in the charge state of movement due to the emission or to the  absorption of electromagnetic field, that is, to the charge-field interaction process.  Therefore, in this short distance limit eq. (\ref{dlc}) also carries dynamical information, not only kinematical, as is the case of eqs. (\ref{0}) and (\ref{l}).

It is instructive to have a close look on the physical meaning of eqs. (\ref{xz}) and (\ref{dlc}) for the case of an emitted field.
Eq. (\ref{xz}) is a restriction on the propagation of a single object, the field emitted by the charge at $z(\tau)$, whereas the equation (\ref{dlc})  connects restrictions on the propagation of two distinct physical objects, the electron and its field: $d\tau$ describes a displacement of the electron on its worldline while $d x$ is the four-vector separation between two other points where the electron self-field is being considered. If $d\tau=0$ then $dx$ is lightlike and collinear to K, as $K.dx=0.$  Thus, $dx$ is  related to a same electromagnetic signal at two distinct times. The electromagnetic field at $x+dx$ can be seen as the same field at $x$ that has propagated to there with the speed of light. On the other hand, if $d\tau\not=0$ then $dx$ is not collinear to K and it is related to two distinct electromagnetic signals, emitted at distinct times. See the Figure 2. In this case, the field at $x+dx$ cannot be seen as the same field at $x$ that has propagated to there. It is another field emitted by the charge at another time. This apparently obvious interpretation of the constraint (\ref{dlc}) reveals, however, deep physical implications as it perceives as being distinct objects the fields $F$ in two events that are not along a same $K$. This comes from extended causality requiring a $F$ defined with support on a $K$-light-cone generator and conflicting with local causality in the definition of $A(x)$.

\vglue-1.0cm
\begin{figure}
\vglue-6.0cm

{\hglue3.5cm
\epsfxsize=400pt
\epsfbox{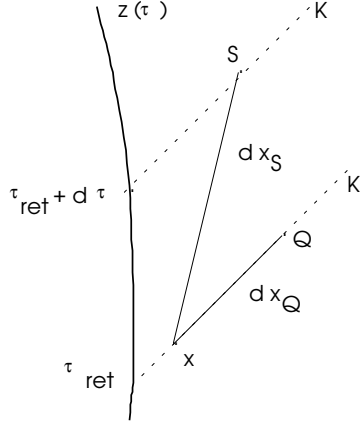}}
\vglue-5.5cm
\caption{The field at the point Q may be considered as the same field at $x$ that has propagated to Q, because $dx_{Q}$ is collinear to K. The fields at events $x$ and S are two distinct signals emitted by the charge at two distinct times $\tau_{ret}$ and $\tau_{ret}+d\tau\;$ as $\;dx_{S}$ is not collinear to K.} 
\end{figure}

But a $F$ defined with  support on a lightcone generator produces strong and experimentally observable consequences. During the free propagation of an electromagnetic radiation, the four-vector $K$ of its light-cone-generator support must be constant. So, the eq. (\ref{dlc}) implies on 
\be
\l{fv}
1+K.u=0
\ee
 and then
\be
\l{dA0}
K.a=0.
\ee
The first one may be seen as a covariant normalization of $K$, that in the charge instantaneous rest frame must satisfy $$K^{0}{\Big|}_{{\vec u}=0}=|{\vec K}|{\Big|}_{{\vec u}=0}=1.$$ The second one is a dynamical constraint between the direction $K$ along which the signal is emitted (absorbed) and the instantaneous change in the charge state of motion at the retarded (advanced) time. 
It implies on
\be
\l{a4}
a_{0}=\frac{{\vec a}.{\vec K}}{K_{0}},
\ee
whereas $a.u\equiv0$ leads to
$a_{0}=\frac{{\vec a}.{\vec u}}{u_{0}},$
and so, in the charge instantaneous rest frame at the limiting emission (absorption) time ${\vec a}$ and ${\vec K}$ are orthogonal vectors,
\be
\l{af}
{\vec a}.{\vec K}{\Big |}_{{\vec u}=0}=0.
\ee
This is an observable consequence of extended causality. For the electromagnetic field this is an old well known and experimentally confirmed fact\cite{Jacksonch14,Rorhlichp112,Ternov}.  Its experimental confirmation is a direct validation of extended causality and of its implications, as discussed in the Section X of \cite{hep-th/9911233}.
 The constraint (\ref{dA0}) that takes, in the standard formalism of continuous fields, the whole apparatus of Maxwell's theory to be demonstrated \cite{Rorhlichp112} can been obtained  on very generic grounds of causality\cite{hep-th/0006237} without reference to any specific interaction. This makes of it a universal relation, supposedly valid for all kinds of fields and sources. This same behaviour, expressed in eq. (\ref{af}), is then expected to hold for all fundamental (strong, weak, electromagnetic and gravitational) interactions.
 
The relevance of eq. (\ref{dA0}) is on its focus on the charge-field interaction process. It is strongly dependent on $K$ being taken as a constant during the field propagation.
A non-constant $K$ would imply on a continuing interaction and this would change the above results. 
 \be
\l{fmn}
\partial_{\mu}K_{\nu}=\frac{1}{\rho}(\eta_{\mu\nu}+K_{\mu}u_{\nu}+K_{\nu}u_{\mu}-K_{\mu}K_{\nu})-K_{\mu}K_{\nu}a.K=\partial_{\nu}K_{\mu}:=K_{\mu\nu},
\ee
from eqs. (\ref{K}) and (\ref{delp}). Then the hypothesis of a non-constant $K$ would not affect eq. (\ref{fv}) because 
\be
K_{\mu\nu}\Delta x^{\nu}=\rho K_{\mu\nu}K^{\nu}=K_{\mu}(1+K.u)\equiv0,
\ee
but 
eq. (\ref{dA0}) would be replaced by just an identity as
\be
\nabla_{\mu}(1+K_{\nu}u^{\nu})=K_{\mu\nu}u^{\nu}-K_{\mu}K_{\nu}a^{\nu}=K_{\mu}(1+K.u)\equiv0.
\ee
So, it is clear that the validity of eq. (\ref{dA0}) rests on a free propagation of the field right after its emission (or, symmetrically, right before its absorption) which indicates no self-interaction, a definitive detachment of the field from its source. Self interaction for the emited field would also imply, by symmetry, causality violation for the absorbed field as it would be interacting with the charge even before reaching it.

\section{The double limit $x\rightarrow z(\tau)$}

The above conclusions can be made more evident considering  the fate of both eqs. (\ref{l}) and (\ref{dlc}) in the limit when the event $x$ approaches the event $z(\tau_{s})$ and its implications to the field energy-tensor\footnote{This is discussed in \cite{hep-th/9610028} but for completeness, considering its relevance here, its main steps and some further considerations are aligned.}.    Nothing obviously happens to the first one; $\Delta x$ just goes to zero. To the second one the restriction connecting $d\tau$ to $dx$ becomes indeterminated because K is not defined in this limit:
\be
\l{i}
\lim_{x\to{z(\tau_{s})}}K=\lim_{x\to{z(\tau_{s})}}\frac{\Delta x}{-u.\Delta x}=\frac{0}{0}?
\ee 
For a lightlike signal, eqs. (\ref{l}) and (\ref{dlc}) together require that the pair of events $x$ and $z(\tau)$ belongs to a same lightcone generator, so that eq. (\ref{i}) can be written  as
\be
\l{li} 
\lim_{x\to{z(\tau_{s})}}K{\bigg|}_{\Delta x^2=0\atop{d\tau+K.dx=0}}=\frac{0}{0}?
\ee
 This notation intends to denote that $x$ approaches $z(\tau_{s})$ through a K-light-cone generator, i.e. by the straight line intersection of the hypercone $(\Delta x^{2}=0$) and its tangent hyperplane ($d\tau+K.dx=0$), eliminating any ambiguity in the definition of the limit in eq. (\ref{i}). Now one can apply the L'H\^opital's rule for evaluating K on the neighbouring events of $z(\tau_{s})$ along the electron worldline, i.e., at either $\tau_{s}+ d\tau$ or $\tau_{s}- d\tau.$ This corresponds to replacing the above simple limit of $x\rightarrow z(\tau_{s})$ by a double and simultaneous  limit of $x\rightarrow z(\tau)$ along the K-lightcone generator while $z(\tau)\rightarrow z(\tau_{s})$ along the electron worldline. This simultaneous double limit is pictorially best described by the sequence of points S, Q,...,P in the Figure 3; each point in this sequence belongs to a K-generator of a lightcone with vertex at the electron worldline $z(\tau).$ 
\begin{figure}
\vglue-6cm

{\hglue3.5cm
\epsfxsize=400pt
\epsfbox{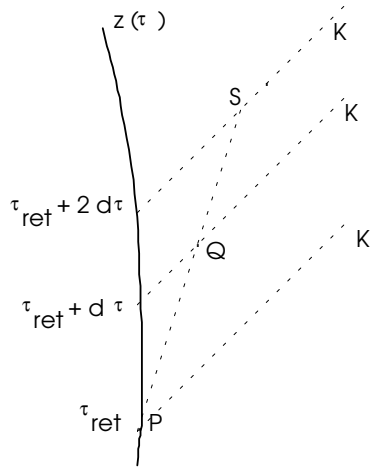}}
\vglue-5cm
\caption{
 The double limit $x\rightarrow z(\tau_{ret})$ along the SQ...P line consists of $x\rightarrow z(\tau)$ along the light-cone generator K while $\tau\rightarrow\tau_{ret}$ on the electron worldline.}
\end{figure}
Then, from eq. (\ref{i}),
$$
\lim_{x\to{z(\tau)}\atop\tau\to\tau_{s}}K{\Big|}_{\Delta x^2=0\atop{d\tau+K.dx=0}}=\lim_{x\to{z(\tau)}\atop\tau\to\tau_{s}}\frac{\Delta{\dot  x}}{-a.\Delta x+u.\Delta{\dot x})}{\bigg|}_{\Delta x^2=0\atop{d\tau+K.dx=0}}=$$
\be
\l{ll}
=\lim_{x\to{z(\tau)}\atop\tau\to\tau_{s}}\frac{{-u}}{u.u}{\bigg|}_{\Delta x^2=0\atop{d\tau+K.dx=0}}
=u,
\ee

as ${\dot \Delta x}:=\frac{d\Delta x}{d\tau}=-u$ and $u^{2}=-1.$ So $K{\big|}_{x=z(\tau_{s})}$ is indefinite but $K{\big|}_{x=z(\tau_{s}\pm d\tau)}=u.$  

The lightlike four-vector K is replaced by the timelike four-vector u in the above defined (double) limit of $\Delta x\rightarrow0.$ This result changes the usual vision of field theory in this limit.

The electron self-field energy tensor, $4\pi\Theta=F.F-\frac{\eta}{4}F^{2}$, after eq. (\ref{nablaA}) becomes,
\begin{equation}
\label{t'}
-4\pi\rho^{4}\Theta=(KW+WK)+KKW^{2}+WWK^{2}+\frac{\eta}{2}(1-K^{2}W^{2}), 
\end{equation}
as $K.u=-1$ from eq. (\ref{K}) and $K.W=-1$. The eq. (\ref{t'}), like eqs. (\ref{aA}) and (\ref{F}),  are all constrained by eq. (\ref{l}), i.e. by $\tau=\tau_{s},$ and they are valid only for $\rho\not=0$, region where $K^{2}=0.$ So, instead of eq. (\ref{t'})  one may write 
\be
\l{rt}
-4\pi\rho^{4}\Theta{\bigg|}_{K^{2}=0}=(KW+WK)+KKW^{2}+\frac{\eta}{2},\qquad\hbox{ for}\quad \rho>0,\;\tau=\tau_{s},
\ee
which corresponds to the usual expressions that one finds, for example in [2-7]. They are equivalent, as long as $\rho>0$.
The four-vector momentum associated to the electron self-field is defined by the flux of its $\Theta$ through a hypersurface $\sigma$ of normal n:
\be
\l{P}
P=-\int d^{3}\sigma n.\Theta{\bigg|}_{K^{2}=0},
\ee
but  $\Theta$ contains a factor $\frac{1}{(\rho)^{4}}$ and this makes P highly singular at $\rho=0$, that is at $x=z(\tau_{s}).$  This is the old well-known self-energy problem of classical electrodynamics which heralds \cite{JS} similar problems in its quantum version. This divergence at $\rho=0$ is also the origin of nagging problems on finding a classical equation of motion for the electron [2-7]. 
But it is clear now, after equation (\ref{ll}), that the standard practice of replacing everywhere $\Theta$ by $\Theta{\bigg|}_{K^{2}=0}$ is not justified and, more than that, it is the cause of the above divergence problem and the related misconceptions in classical electrodynamics. One must use eq. (\ref{t'}), the complete expression of $\Theta$, in eq. (\ref{P}) and repeat for it the same double limit done in eq. (\ref{ll}). The long but complete and explicit calculation is done in \cite{hep-th/9610028}; its results and conclusions are summarised here:\\ $P{\bigg|}_{x=z(\tau_{s})}$ is undefined but
\be
\l{PO}
P{\bigg|}_{x=z(\tau_{s}-)}=P{\bigg|}_{x=z(\tau_{s}+)}=0.
\ee
There is no infinity at $\rho=0$! This infinity disappears only when the double limiting process is taken because the lightcone generator $K$ must then be recognized as the actual support of the Maxwell field $F.$ The message here is that the infinities and other inconsistencies of classical electrodynamics are not to be blamed on the point electron but on the lightcone support of the field in the eq. (\ref{aA}). Extended causality cannot just be ignored.

\section{Conclusions}

We can summarize it all with the following implications to the charge-field dynamics:
\begin{enumerate}
\item No self interaction; once emitted the field no longer interacts with the charge;
\item The emission process is discrete;
\item The emission event is an isolated singularity on the charge worldline, in the sense of discontinuity on its first derivative.
\end{enumerate}

Equation (\ref{PO}) confirms that $z(\tau_{s})$ is an isolated singularity. This is in direct contradiction to the standard view of a continuous field, emitted or absorbed by the charge in a continuous way. According to eq. (\ref{PO}) there is no charge self field at $z(\tau_{s}\pm d\tau)$, but only sharply at $z(\tau_{s})$.
This is puzzling! It is saying that the Gauss' law, in the zero distance limit, $lim_{S\rightarrow0}\int_{S}d\sigma{\vec E}.{\vec n} = 4\pi  e$, is meaningful only at $z(\tau_{s})$ and not at $z(\tau_{ret\pm d\tau})$ because ${\vec E}(\tau_{s})\not=0$ but ${\vec E}(\tau_{ret\pm d\tau})=0.$\\
It implies, in other words, that the electromagnetic interactions are discrete and localized in time and in space. In terms of a discrete field interaction along a lightcone generator, as the one represented\footnote{For simplicity, the cause(the absorption of a photon, for example) of the sudden change in the electron state of movement is not shown, only its consequence (the emission of a photon)} in the Figure 4, one can understand the physical meaning of eqs. (\ref{li}), (\ref{ll}) and  (\ref{PO}). 
The Maxwell fields are just effective average descriptions of an actually discrete interaction field. The field discreteness (or the existence of photons) is masqueraded by this averaged field and it takes the zero distance limit to be revealed from the Maxwell field. It may sound unbelievable or even suspicious that these conclusions have been derived exclusively from the supposedly exhaustively known classical electrodynamics but nothing has been added to or modified in the old Maxwell's theory, except a new interpretation of old results. They come from the recognition of the existence of two mutually excludent causality-implementation modes in the Maxwell's formalism. This could have been taken, if it were known at the beginning of the last century, as a first indication of the quantum, or of the discrete nature of the electromagnetic interaction. All these are consequences of the dynamical constraints hidden on the restrictions (\ref{l}) and (\ref{dlc}).

The initial goal of pointing and discussing the implicit existence of two distinct modes of implementing causality in field theory and that its consistency problems are created by non-recognizing them has been fulfilled. A completely consistent field formalism must be expressed in term of fields defined ab initio with extended causality. How this can be accomplished, its consequences and how it is related to the standard formalism based on local causality only is done in \cite{hep-th/0006237}.

\begin{figure}
\hglue2.0cm 
\epsfxsize=400pt
\epsfbox{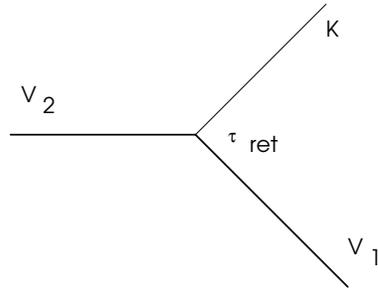}
\vglue-12cm
\caption{
A discrete interaction along a lightcone generator $K.$ There is no electron self-field immediately before or after $\tau_{ret}.$ It is an isolated singularity. $\tau_{ret}$ is a singular point on the electron worldline only because its tangent is not defined there; there is no infinity.}
\end{figure}

\end{document}